\documentclass[11pt]{article}
\usepackage{iap2000,epsfig}
\usepackage{graphicx}
\usepackage[dvips]{color}

\def\Journal#1#2#3#4{{#1} {\bf #2}, #3 (#4)}

\newfont{\winzig}{cmr5 scaled 1000}
\newfont{\klein}{cmr12 scaled 1000}
\newfont{\kleinfett}{cmr12 scaled 1000}

\newcommand{\dd}{{\rm d}}

\newcommand{\V}[1]{\vec{#1}}

\newcommand{\Wab}{ W(\V{r}_a-\V{r}_b,h) }

\begin{document}
\title{PROPERTIES OF SIMULATED MAGNETIZED GALAXY CLUSTERS
\\[16pt]
\mbox{\includegraphics[width=0.35\textwidth]{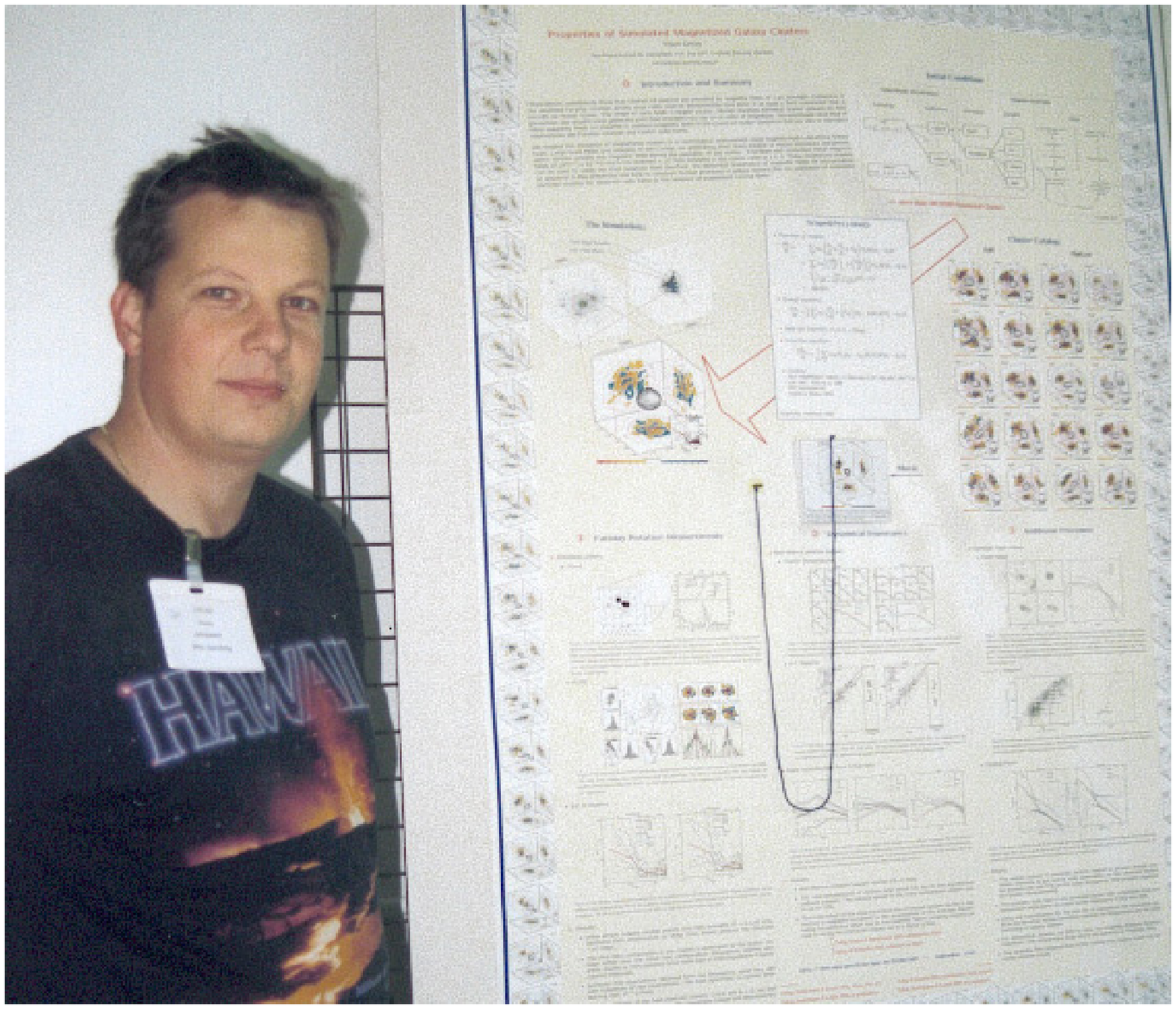}
}}

\author{ Klaus Dolag }

\address{Max-Planck-Institut f\"ur Astrophysik, P.O.~Box
       1317, D--85741 Garching, Germany}

\maketitle\abstracts{
We study the evolution of magnetized clusters in a cosmological
environment using magneto-hydro dynamical simulations. Large scale
flows and merging of subclumps generate shear flows leading to 
Kelvin-Helmholtz instabilities, which, in addition to the compression
of the gas where the magnetic field is frozen in, further amplify the magnetic 
field during the evolution of the cluster. Therefore, well-motivated 
initial magnetic fields of $\langle B^2\rangle^{1/2}=10^{-9}\,{\rm G}$ 
reach the observed $\sim\mu{\rm G}$ field strengths in the cluster 
cores at $z=0$. These magnetized clusters can be used to study the final magnetic
field structure, the dynamical importance of magnetic fields for
the interpretation of observed X-Ray properties, and help to constrain
further processes in galaxy clusters like the population of
relativistic particles giving rise to the observed radio halos or the
behavior of magnetized cooling flows.
}

\section{Introduction}
Observations consistently show
that clusters of galaxies are pervaded by magnetic fields of
$\sim\mu{\rm G}$ strength.  Coherence of the observed Faraday rotation
across large radio sources demonstrates that there is at least a field
component that is smooth on cluster scales.
The origin of such fields is largely unclear. Models invoking
individual cluster galaxies for field generation and amplification
generally yield field strengths too low by an order of magnitude.

We used the cosmological MHD code described in Dolag et 
al. (1999) to simulate the formation of magnetised galaxy clusters 
from an initial density perturbation field. Our main results can be
summarised as follows: (i) Initial magnetic field strengths are
amplified by approximately three orders of magnitude in cluster cores,
one order of magnitude above the expectation from flux conservation
and spherical collapse. (ii) Vastly different initial field configurations
(homogeneous or chaotic) yield results that cannot significantly be
distinguished. (iii) Micro-Gauss fields and Faraday-rotation
observations are well reproduced in our simulations starting from
initial magnetic fields of $\sim10^{-9}\,{\rm G}$ strength at redshift
15. Our results show that (i) shear flows in clusters are crucial for
amplifying magnetic fields beyond simple compression, (ii) final field
configurations in clusters are dominated by the cluster collapse
rather than by the initial configuration, and (iii) initial magnetic
fields of order $10^{-9}\,{\rm G}$ are required to match
Faraday-rotation observations in real clusters.

We used these magnetized clusters to study the final magnetic
field structure, the dynamical importance of magnetic fields for
interpretation of observed X-Ray properties and start to constrain
further processes in galaxy clusters like the population of
relativistic particles causing the observed radio halos or the
behavior of magnetized cooling flows.
 
\section{GrapeSPH+MHD}
The code combines the merely gravitational interaction of a
dark-matter component with the hydrodynamics of a gaseous
component. The gravitational interaction of the particles is evaluated
on GRAPE boards (Sugimoto et al. 1990), while the gas dynamics is computed in the SPH
approximation. It was also supplemented with the magneto-hydrodynamic
equations to trace the evolution of the magnetic fields which are
frozen into the motion of the gas because of its assumed ideal
electric conductivity. The back-reaction of the magnetic field on the
gas is included. It is based on GrapeSPH (Steinmetz 1996) and solves the
following equations:
   \begin{itemize}      
      \item[{\color{red}$\bullet$}] {\color{blue} Equation of motion:}
\begin{eqnarray}
   \frac{\dd\V{v}_a}{\dd t}= &-& \sum_b m_b\left(\frac{P_b}{\rho_b^2}+
                                 \frac{P_a}{\rho_a^2}+\Pi_{ab}
                                 \right)\nabla_a\Wab \nonumber \\
                             &+& \sum_b m_b\left[\left(\frac{{\cal M}_{ij}}{\rho^2}\right)_a+
                                 \left(\frac{{\cal M}_{ij}}{\rho^2}\right)_b
                                 \right]\nabla_{a,j}\Wab \nonumber \\
                             &-& \sum_i\frac{m_i}{(|\V{r}_a-\V{r}_i|^2+\epsilon_a^2)^{1.5}}
                                 (\V{r}_a-\V{r}_i) \nonumber \\
                             &+& \Omega_\Lambda^0H_0^2\V{r}_a \nonumber
\end{eqnarray}
      \item[{\color{red}$\bullet$}] {\color{blue} Energy equation:}
$$
   \frac{\dd u_a}{\dd t} = \frac{1}{2}\sum_b m_b\left(
                           \frac{P_a}{\rho_a^2}+ \frac{1}{2}\Pi_{ab}
                           \right)(\V{v}_a-\V{v}_b)\nabla_a\Wab
$$
      \item[{\color{red}$\bullet$}] {\color{blue} Ideal gas equation:}
      $ P_i=(\gamma-1)u_i\rho_i. $
      \item[{\color{red}$\bullet$}] {\color{blue} Induction equation:}
$$
   \frac{\dd \V{B}_{a,j}}{\dd t}=\frac{1}{\rho_a}\sum_b m_b(
             \V{B}_{a,j}\V{v}_{ab}-\V{v}_{ab,j}\V{B}_a)\nabla_a\Wab
$$
      \item[{\color{red}$\bullet$}] {\color{blue} Cooling:} \\
          Non-equilibrium solver, 6 Species H,H$^+$, He, He$^+$,
          He$^{++}$, e$^-$
          (Cen 1992 , Katz et al. 1996)
      \item[{\color{red}$\bullet$}] {\color{blue} Heating:} \\
          UV background (Haardt \& Madau 1996) \\

   \end{itemize}

\section{Initial Conditions}
\begin{figure}[ht]
\begin{center}
\psfig{figure=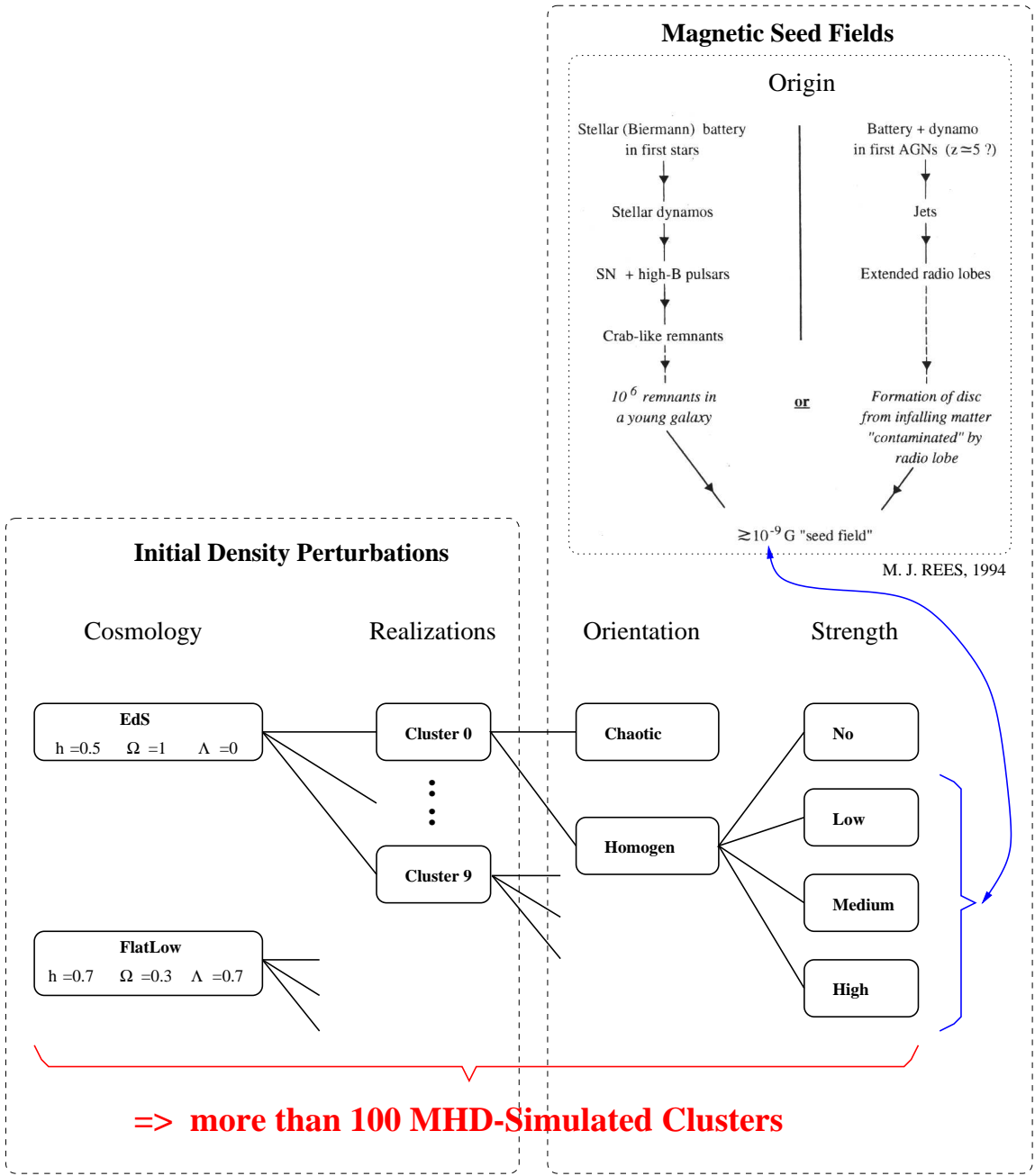,width=0.95\textwidth}
\end{center}
\caption{A Schematic overview of the initial conditions is given.
Useful combinations of them, for the seed magnetic field as well as for the initial 
density fluctuations, lead to more than 100 necessary simulations.
\label{fig:ic}}
\end{figure}

We need two types of initial conditions for our simulations, namely
(i) the cosmological parameters and initial density perturbations, and
(ii) the properties of the magnetic seed field.
Two different kinds of cosmological models are used, EdS and FlatLow.
For each cosmology, we calculate ten different realisations which result in clusters 
of different final masses and different dynamical states at redshift 
$z=0$. We simulate each of these clusters with 
different initial magnetic fields, yielding a total of more than 100
cluster models. Since the origin of magnetic fields on cluster scales 
is unknown, we use either completely homogeneous or chaotic 
initial magnetic field structures. An overview of the initial
conditions is given in Figure \ref{fig:ic}.

\begin{figure}[ht]
\begin{center}
\psfig{figure=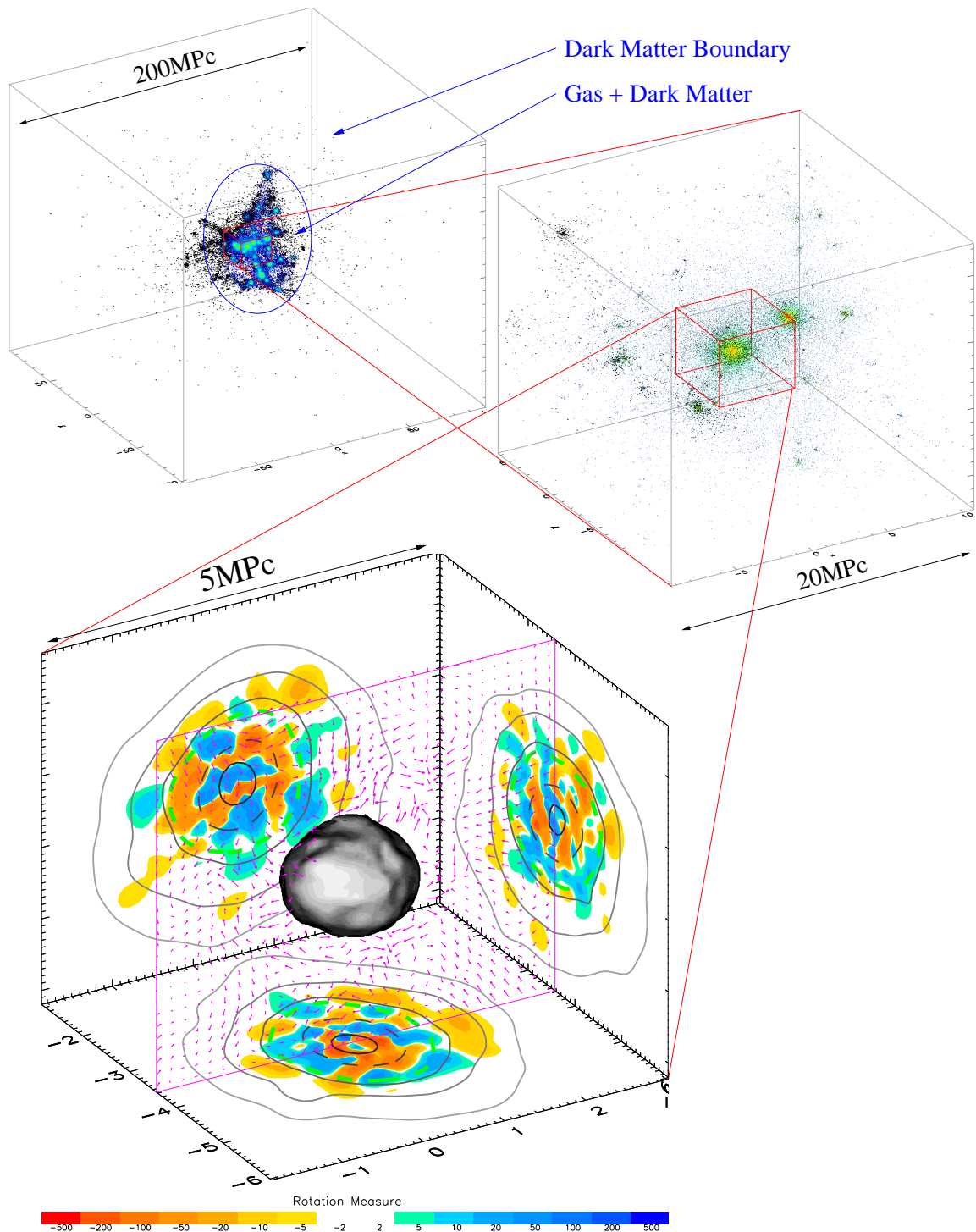,width=0.95\textwidth}
\end{center}
\caption{Shown is the particle distribution inside a simulation box,
where dark matter particles are black, and the gas particles are colored
according to their density.
The lower box in this figure shows one simulated cluster in a
three-dimensional visualisation.
The Faraday-rotation measures produced by the
cluster in the three independent spatial directions are projected onto
the box sides and encoded by the color scale as indicated below the
box. The gray solid curves are projected density contours, whereas
the dashed line marks half the central density. 
The green dashed curve encompasses the region emitting
90\% of the projected X-ray luminosity. The shaded object in the
center is the isodensity surface at hundred times the critical density. 
In addition, magnetic field vectors are plotted in the inserted slice 
marked by the purple rectangle. Coordinates are physical coordinates in Mpc.
\label{fig:sim}}
\end{figure}

\section{The Simulation}
\begin{figure}[ht]
\begin{center}
\psfig{figure=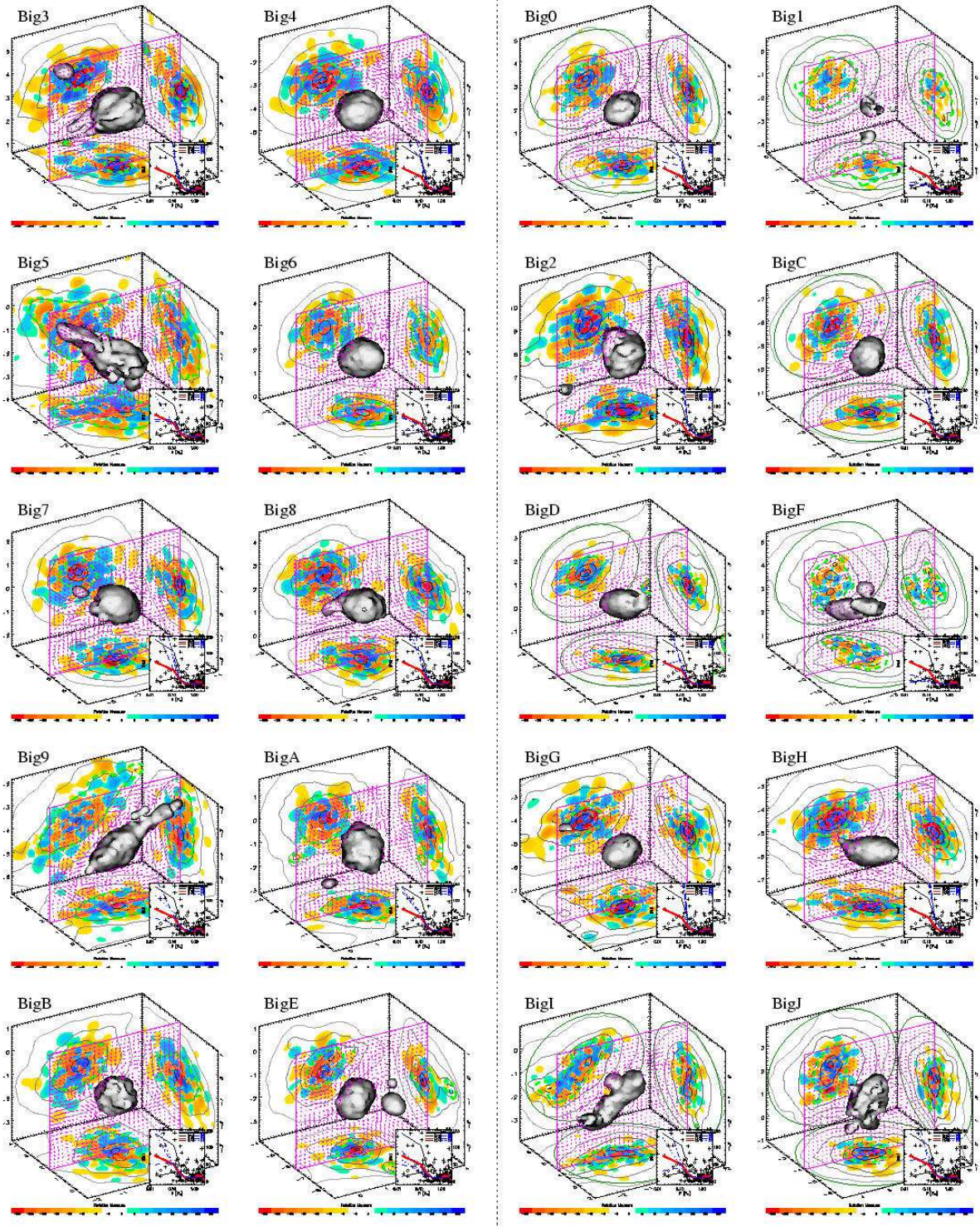,width=0.99\textwidth}
\end{center}
\caption{Visualization of all ten clusters in both cosmologies. Shown
are the same quantities as in the lower panel of figure \ref{fig:sim}.
\label{fig:cat}}
\end{figure}

The simulations consist of a dissipation-free dark matter component
interacting only through gravity, and a dissipational, gaseous
component. The surroundings of the clusters are dynamically important
because of tidal forces and the details of the merger history.
To account for that, the cluster simulation volumes are
surrounded by a layer of boundary particles which accurately
represent accurately the sources of the tidal fields 
in the cluster neighborhood. Figure \ref{fig:sim} shows
the structure of one of our simulations, figure \ref{fig:cat} shows
the whole clusters catalog.

\newpage

\section{Results}
\subsection{Faraday Rotation Measurements}
\begin{figure}[ht]
\begin{center}
\psfig{figure=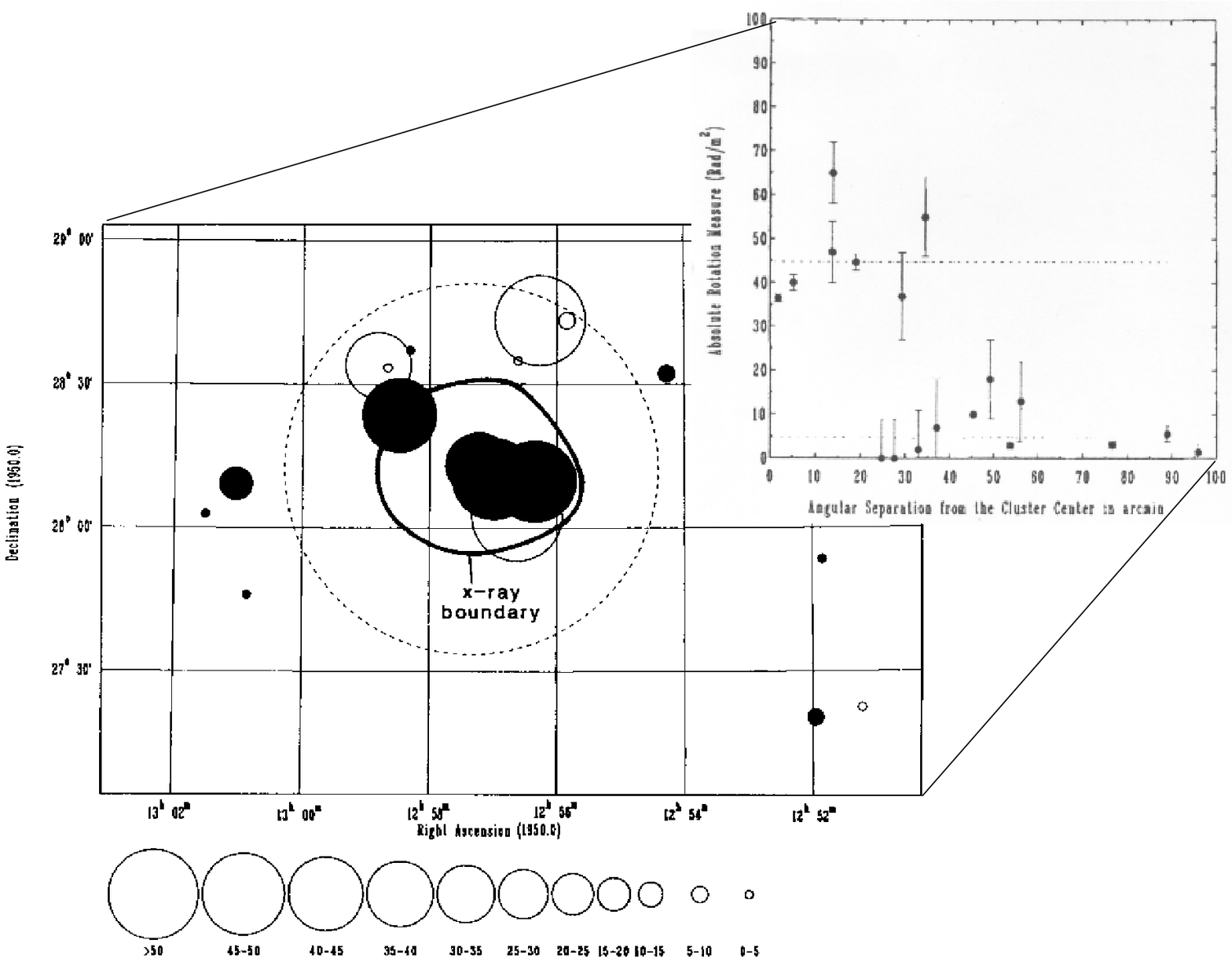,width=0.5\textwidth}
\psfig{figure=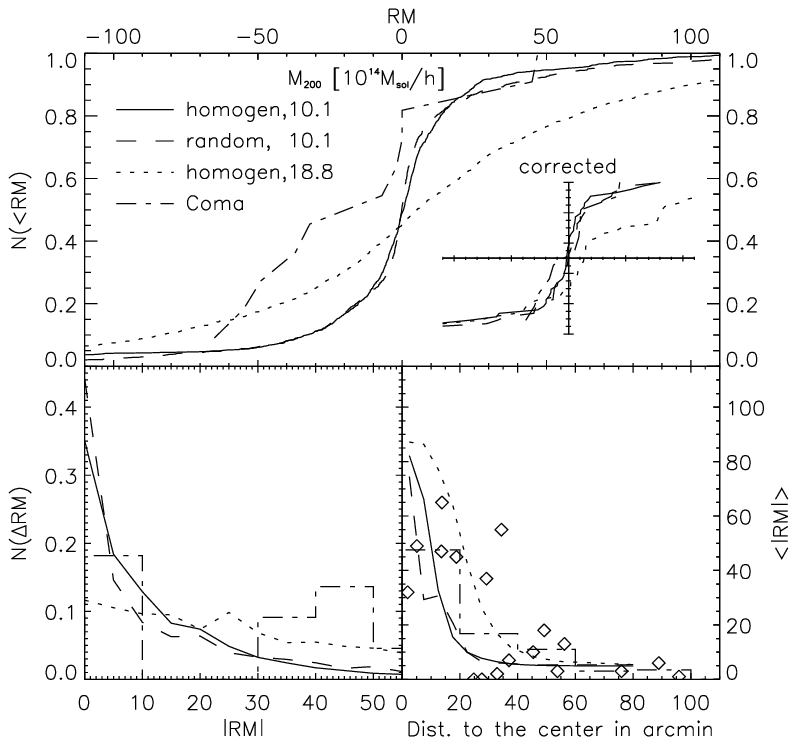,width=0.4\textwidth}
\end{center}
\caption{In the left panel, the circles indicate positions and values of the
observed Faraday-rotation measurements in the Coma cluster (Kim et al.~1990).
Plotted as function of clustercentric distances, the signal
induced by the ICM is clearly visible.
The right panel compares statistically the observations and the
simulations in three different ways. It is obvious, that while 
different simulated clusters give different signals (solid and dotted
lines), one of them matches the observations (dashed-dotted line) 
very well (a Kolmogorov-Smirnov test gives a 30\% probability for 
the observation to be drawn from the
simulated distributions for the smaller cluster).
\label{fig:coma}}
\end{figure}

\begin{figure}[b]
\begin{center}
\psfig{figure=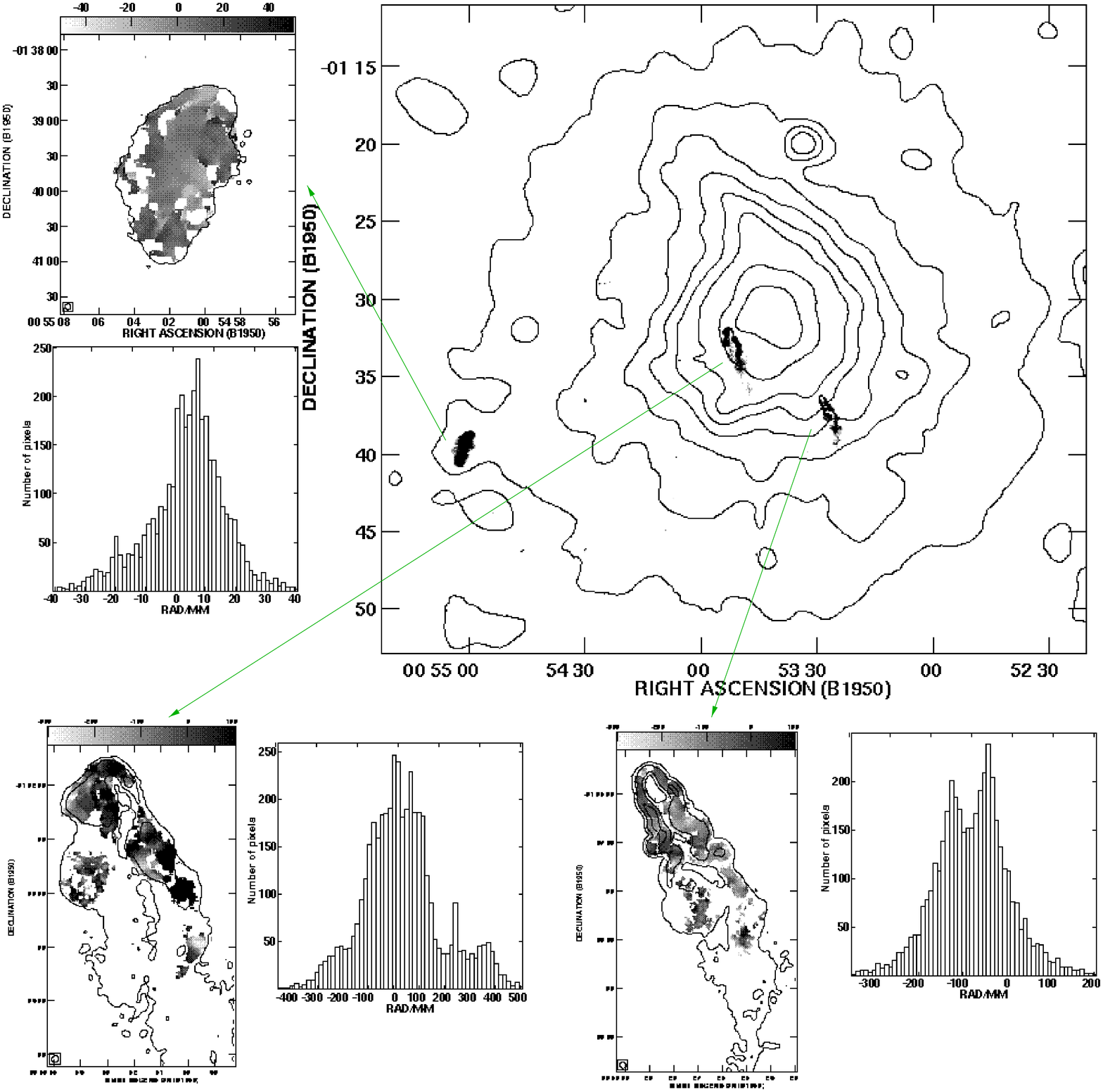,width=0.5\textwidth}
\psfig{figure=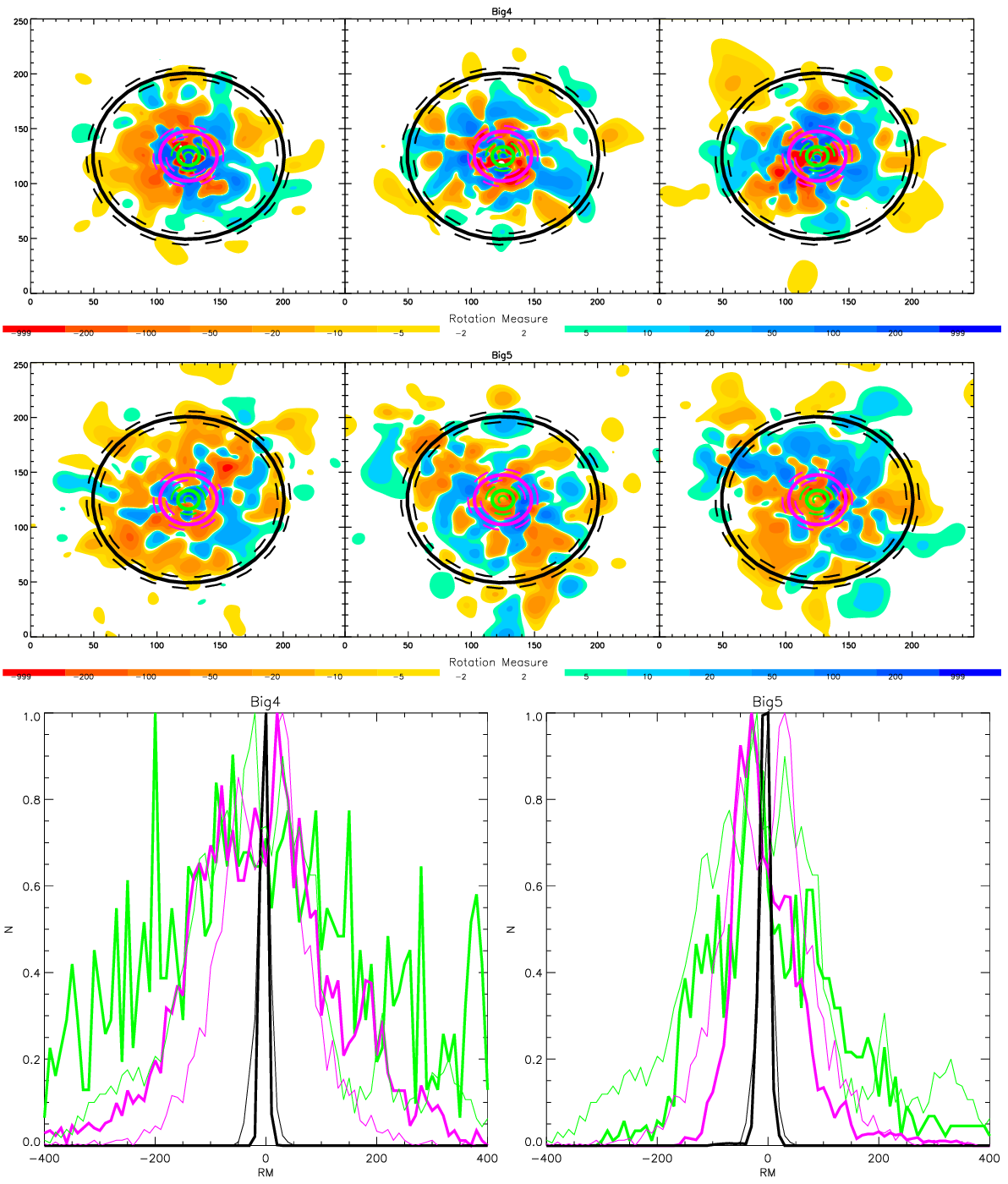,width=0.4\textwidth}
\end{center}
\caption{The left panel shows the RM distribution measured from three
elongated radio sources in A119 (Feretti et al.~1999). The right
panel compares these measurements with the two simulations Big4 and Big5. The
circles in the synthetic RM maps indicate the position of the
sources. The thick lines in the lower right panels represent the
simulated data, the thin lines are drawn from the observations.
\label{fig:a119}}
\end{figure}

We found that the synthetic Faraday-rotation measurements produced by the 
clusters in our simulations match very well those measured in 
individual clusters like Coma (Figure
\ref{fig:coma}) or A119 (Figure \ref{fig:a119}). For details, see Dolag
et al. 1999.

\begin{figure}[th]
\begin{center}
\psfig{figure=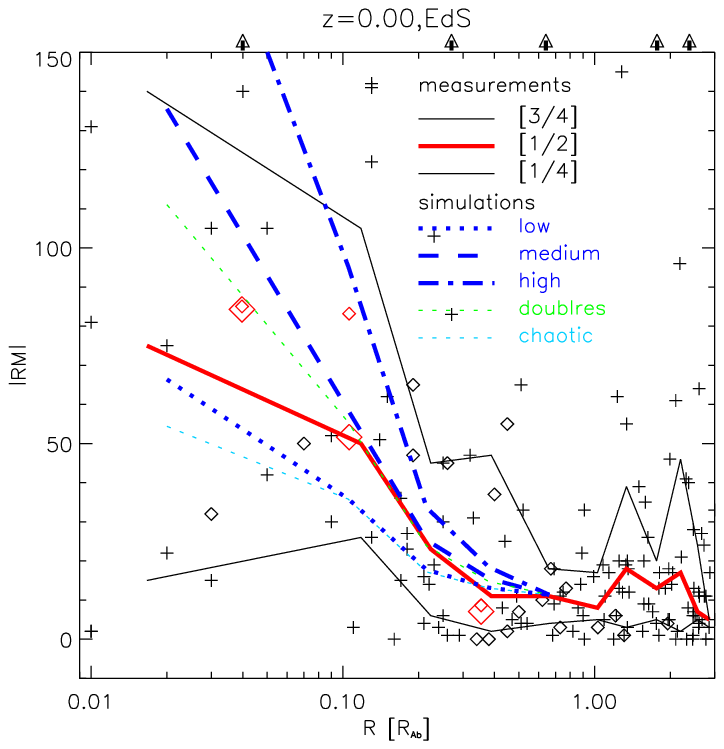,width=0.49\textwidth}
\psfig{figure=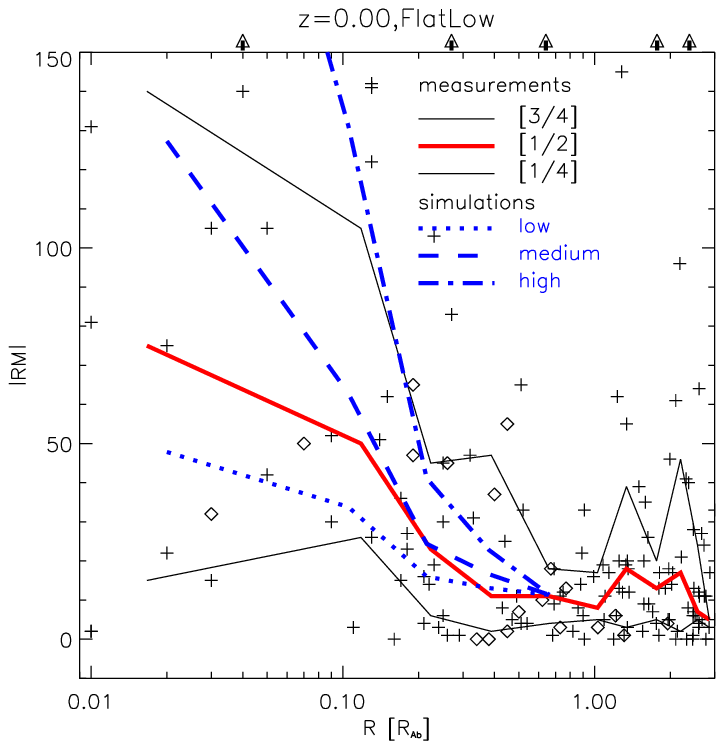,width=0.49\textwidth}
\end{center}
\caption{The Faraday rotation measurements for a set of clusters (Kim et
al.~1991) are compared with the simulations for the different initial
magnetic fields. The red line marks the median of the measurements,
the blue lines are the medians drawn from the simulations. The left 
side shows the results for the EdS, the right side for the FlatLow
cosmology.\label{fig:comp_all}}
\end{figure}

The statistics of the synthetic Faraday-rotation measurements produced by
our simulated cluster sample also match the observations quite well, as
demonstrated in figure \ref{fig:comp_all} for both cosmologies. For
detail see Dolag, Bartelmann \& Lesch (1999, 2000a).
The conclusions drawn from the comparison of synthetic and observed
rotation measurements can be summarized as follows:

   \begin{itemize}
      \item[{\color{blue}$\bullet$}] While simple collapse models
       for motivated initial magnetic fields only predict
        final field strengths of $\sim0.1\,\mu{\rm G}$, additional
        field amplification by shear flows indeed produce the observed
        $\sim\mu{\rm G}$ fields.
      \item[{\color{blue}$\bullet$}] The final field configuration in the
        clusters is dominated by the cluster collapse rather than the
        initial field configuration.
        Simulations starting with either chaotic or homogeneous initial
        fields lead to indistinguishable Faraday rotation measures.
      \item[{\color{blue}$\bullet$}] Synthetic RM observations obtained
          from our simulations agree very well with collections of
          real observations. The best agreement is reached 
          when starting with $\sim10^{-9}\,{\rm G}$ fields at z=15.
      \item[{\color{blue}$\bullet$}] The RM statistic of the 
          best-observed clusters, Coma and A~119, are well reproduced
          by simulated clusters with comparable masses and temperatures.
   \end{itemize}

\newpage

\subsection{Dynamical Importance}
\begin{figure}[th]
\begin{center}
\psfig{figure=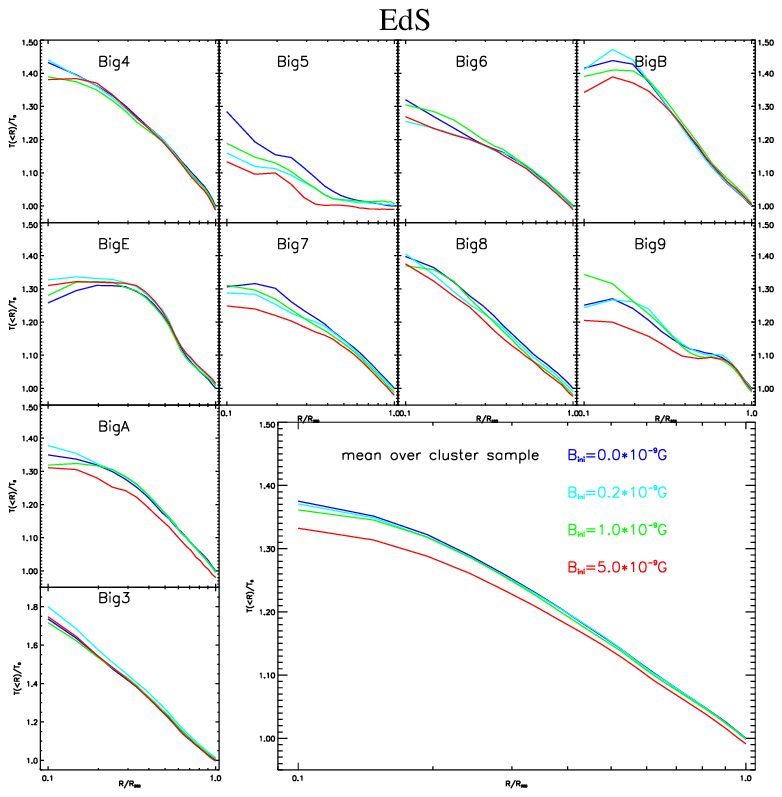,width=0.46\textwidth}
\psfig{figure=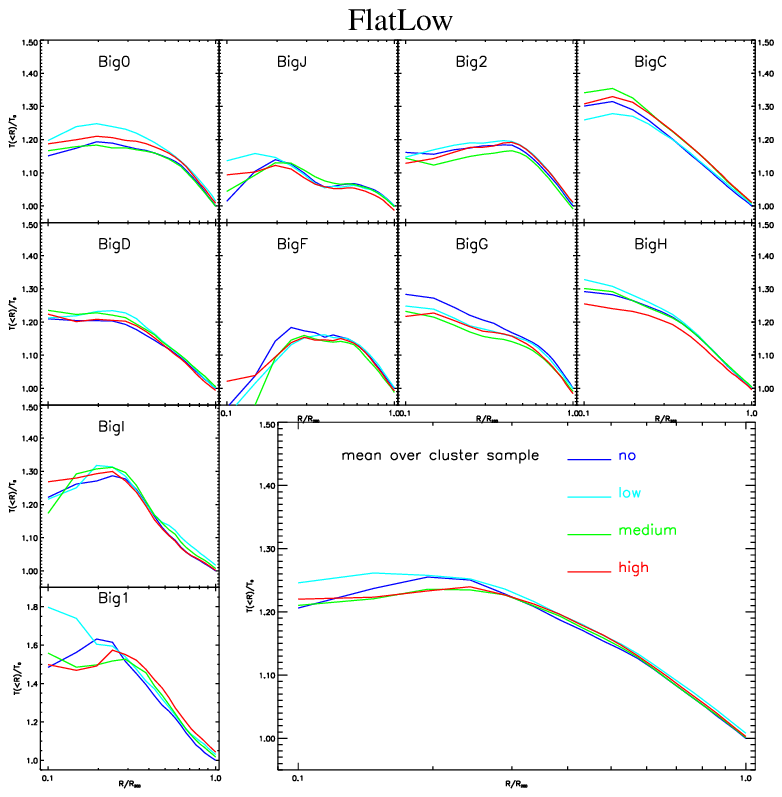,width=0.46\textwidth}
\end{center}
\caption{This figure shows the radial temperature profile in both cosmologies
for each cluster (small panels) and averaged across the simulated
cluster sample (large panels). 
The temperature is scaled by the
mass-weighted temperature of the non-magnetized cluster at the virial radius. 
Different colors distinguish different initial magnetic
field strengths.\label{fig:tprof}}
\end{figure}

\begin{figure}[hb]
\begin{center}
\psfig{figure=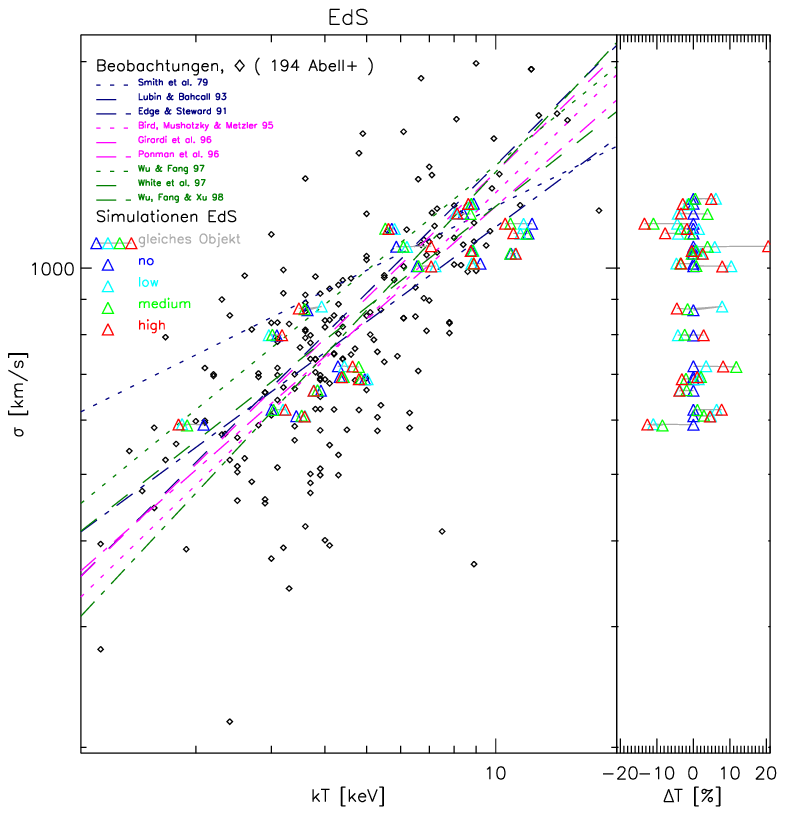,width=0.46\textwidth}
\psfig{figure=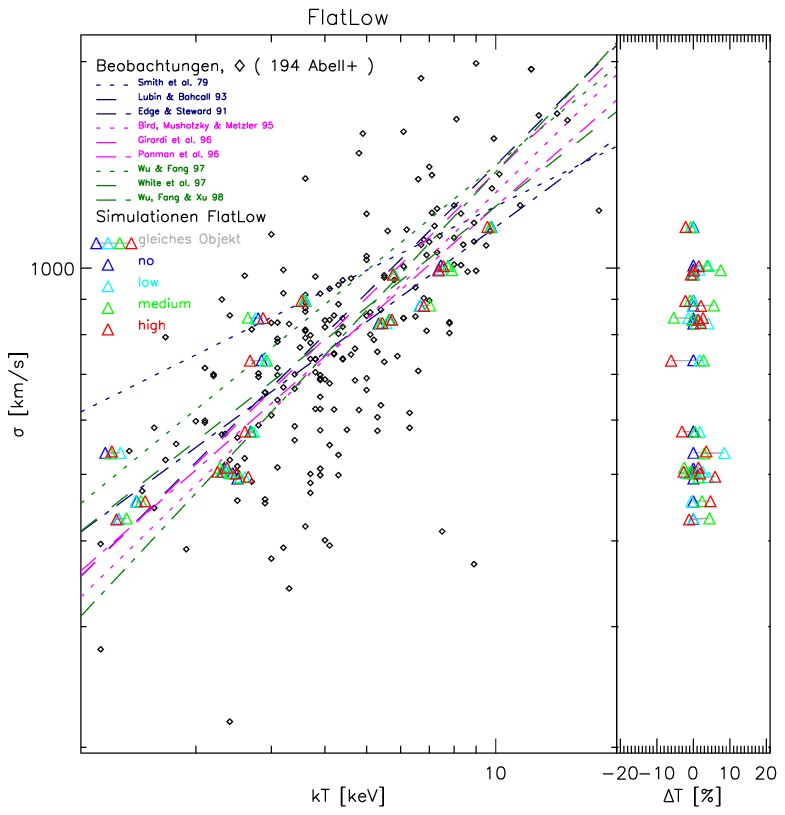,width=0.46\textwidth}
\end{center}
\caption{This figure shows the temperature-mass relation for the
simulated cluster sample. Here, the temperature is the
emission-weighted temperature within the virial radius.
The overlays on the right-hand sides show the temperature change 
in percent for each simulated cluster relative to
non-magnetized clusters for the different initial magnetic field
strengths. \label{fig:tsig}}
\end{figure}

The magnetic fields affect the balance between the
gravitational force and the total (magnetic plus thermal) pressure
in the cluster and therefore can change the temperature of the inter-
cluster medium. Figure \ref{fig:tprof} shows
the change of the temperature in the inter-cluster medium due to the 
presence of magnetic fields in our simulation. Figure
\ref{fig:tsig} shows how this affects the temperature-mass relation in
our simulated cluster samples. For details see Dolag, Evrard \&
Bartelmann (2000).

\begin{figure}[th]
\begin{center}
\psfig{figure=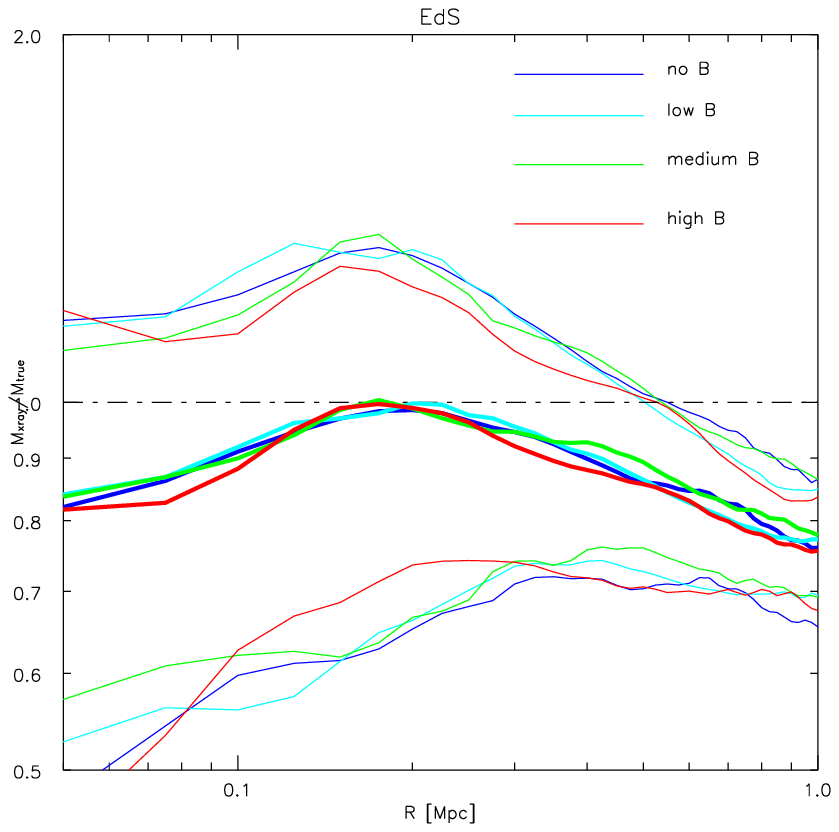,width=0.46\textwidth}
\psfig{figure=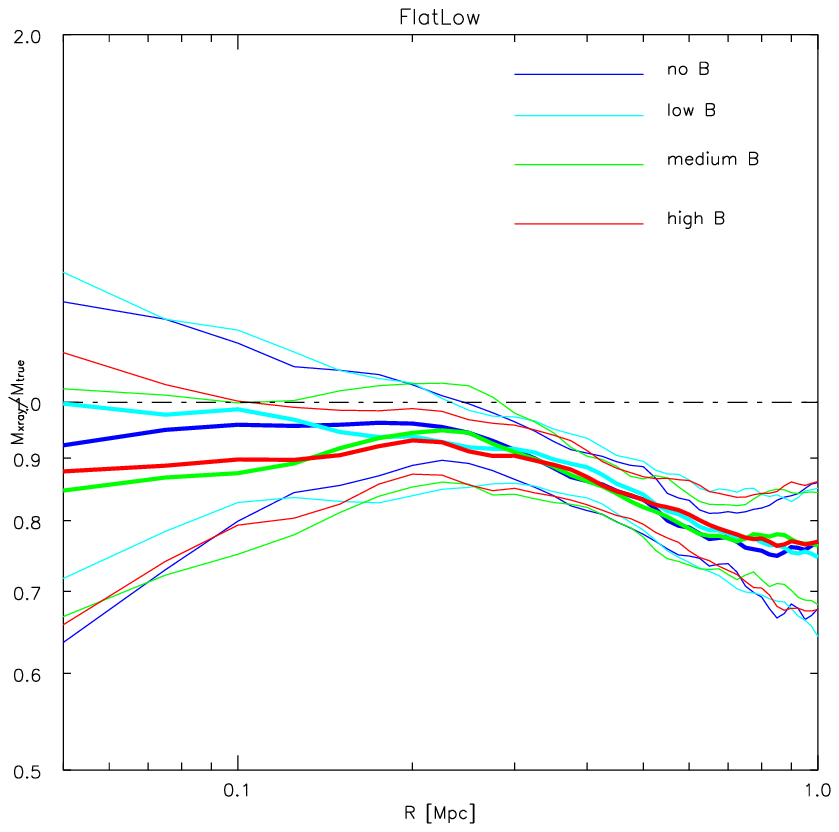,width=0.46\textwidth}
\psfig{figure=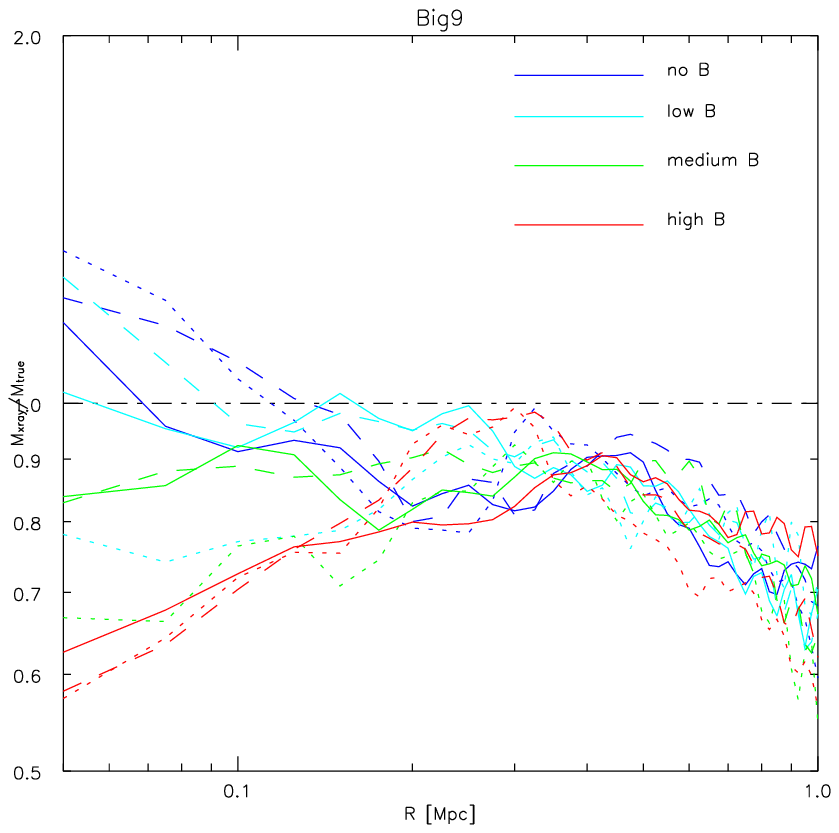,width=0.46\textwidth}

\end{center}
\caption{Ratio of true and X-ray-infered mass for 4 different
magnetic field strengths in two different cosmologies.
The bold lines are averaged profiles of relaxed clusters in the two different
cosmologies, while the thin lines show the corresponding standard deviation.  
The lower panel shows the same for the three projection directions of an individual model.
For this merger the magnetic effects are considerably larger
than the usual scatter of mass profiles from different projection
directions. \label{fig:mxray}}
\end{figure}

The magnetic pressure is not taken into account 
in the X-ray mass-determination methods and therefore potentially 
leads to an underestimation of the mass. Figure \ref{fig:mxray} focusses on the
effect on the mass reconstructed via the X-ray method. For details
see Dolag \& Schindler (2000).
The conclusions drawn from synthetic X-ray observations can be summarized as follows:
   \begin{itemize}
      \item[{\color{blue}$\bullet$}] Non-thermal pressure support 
          reaches 5\% at most.
      \item[{\color{blue}$\bullet$}] The core temperatures of clusters
          drop by about 5\% due to the non-thermal pressure
          support. The induced spread in the mass-temperature relation can be
          up to 15\%. 
      \item[{\color{blue}$\bullet$}] The mean effect on the mass
          reconstruction of relaxed clusters via the X-ray method 
          is negligible compared to the uncertainties of the widely used
          $\beta$-model. Nevertheless, the additional effect due to the
          magnetic field in merging clusters can lead to wrong
          reconstructed masses up to a factor of two. 
   \end{itemize}

\newpage

\subsection{Additional Processes}
\begin{figure}[ht]
\begin{center}
\psfig{figure=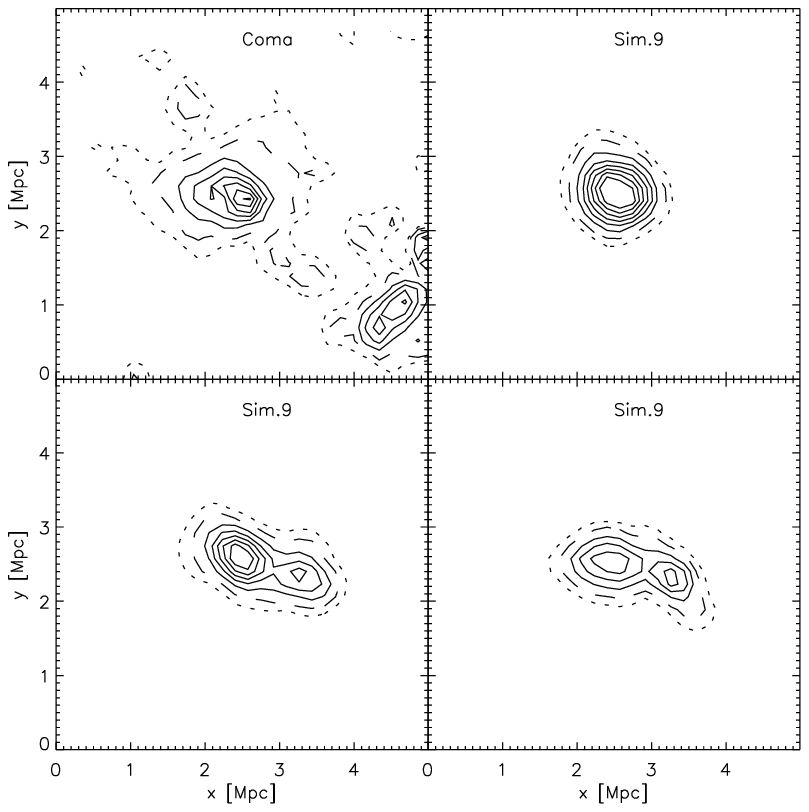,width=0.49\textwidth}
\psfig{figure=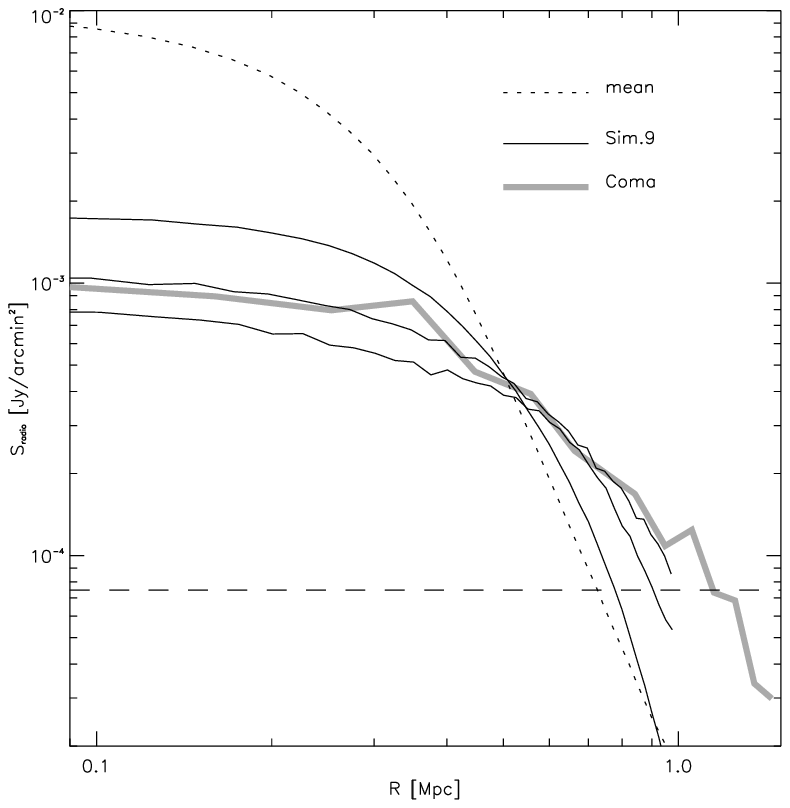,width=0.49\textwidth}
\end{center}
\caption{On the left side, the upper left panel shows the
Effelsberg radio map of the radio halo of the Coma cluster from Deiss
et al. (1997). The other panels show synthetic
radio maps of one simulated cluster (Sim.~9) at 1.4 GHz in three
spatial projections. They are smoothed to the resolution of the radio
observation for comparison. 
The right side compares the radial profile of the Coma cluster (heavy
line) with the profiles obtained from the three projections of the simulation 
Sim.~9 (thin lines). The dotted line represents
the mean profile over all ten simulated clusters. The horizontal
line marks a rough estimate of observational noise. \label{fig:radiomap}}
\end{figure}

\begin{figure}[hb]
\begin{center}
\psfig{figure=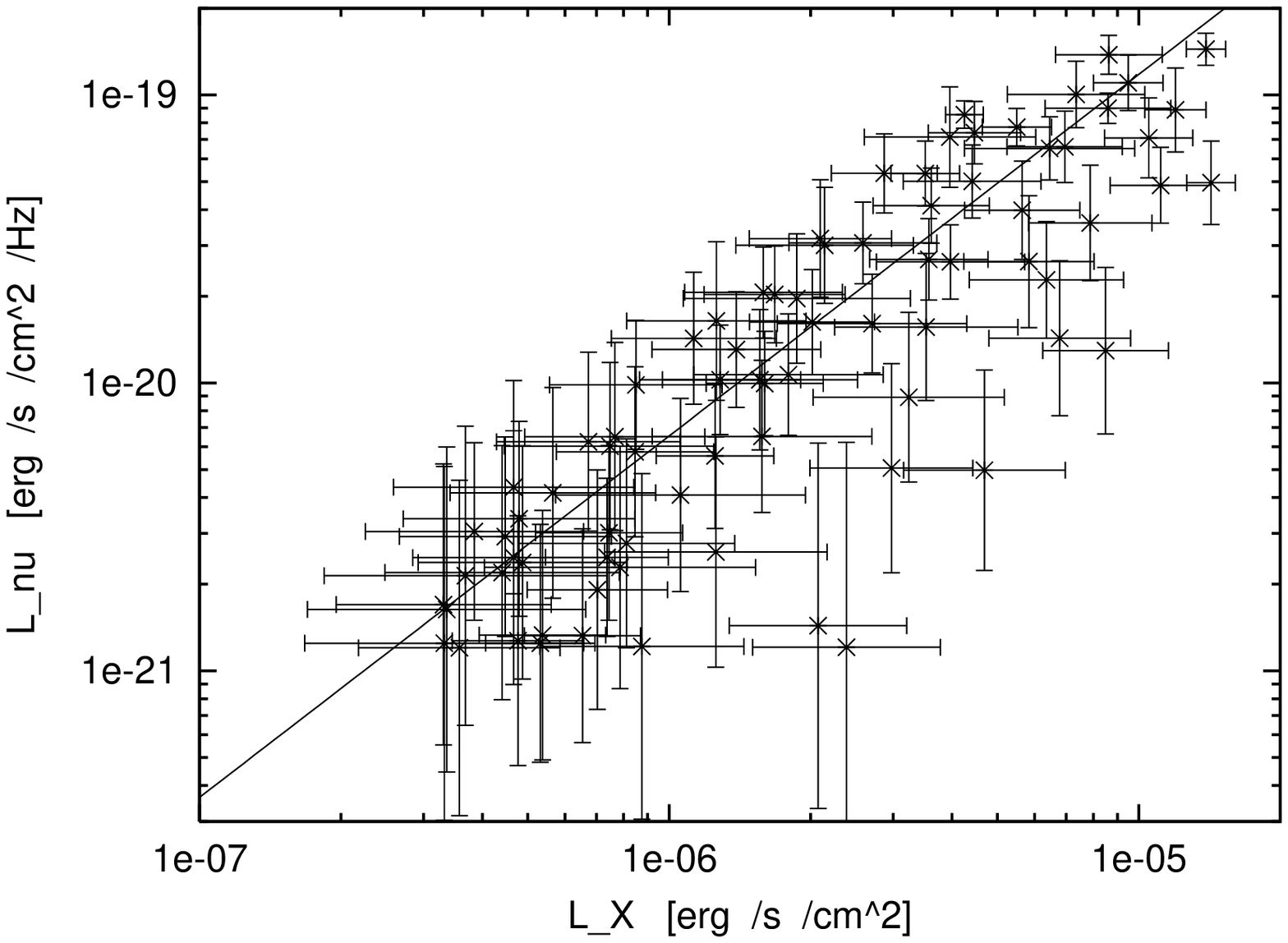,width=0.54\textwidth}
\psfig{figure=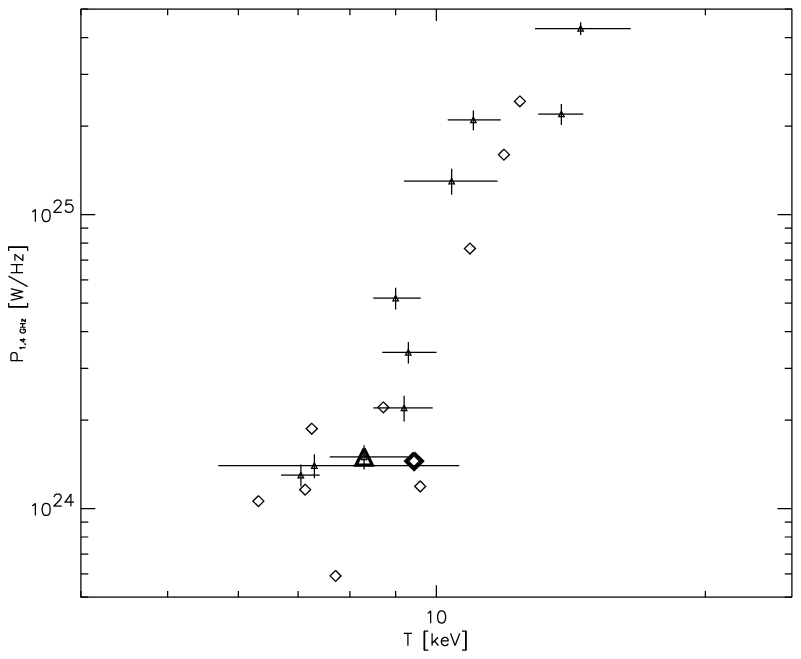,width=0.45\textwidth}
\end{center}
\caption{The left panel shows a point by point radio- to X-ray-emission comparison of
one simulated cluster. The right panel shows the temperature-radio
luminosity relation. The data points with the error
bars are taken from Liang (1999), the diamonds represent the
simulated clusters. Bigger symbols mark Coma and Sim.~9. \label{fig:radiocorr}}
\end{figure}

We demonstrated that a simple model for
hadronic electron injection in a realistic magnetic field
configuration taken from our simulated cluster sample leads to radio halos which reproduce several
types of observations: the profile of the radially decreasing radio emission as
shown in figure \ref{fig:radiomap}, the low radio
polarization, the correlation between radio luminosity an x-ray
surface brightness and
the cluster radio halo luminosity-temperature relation as shown in
figure \ref{fig:radiocorr}. For details see Dolag \& Ensslin (2000).

\newpage

\begin{figure}[ht]
\begin{center}
\psfig{figure=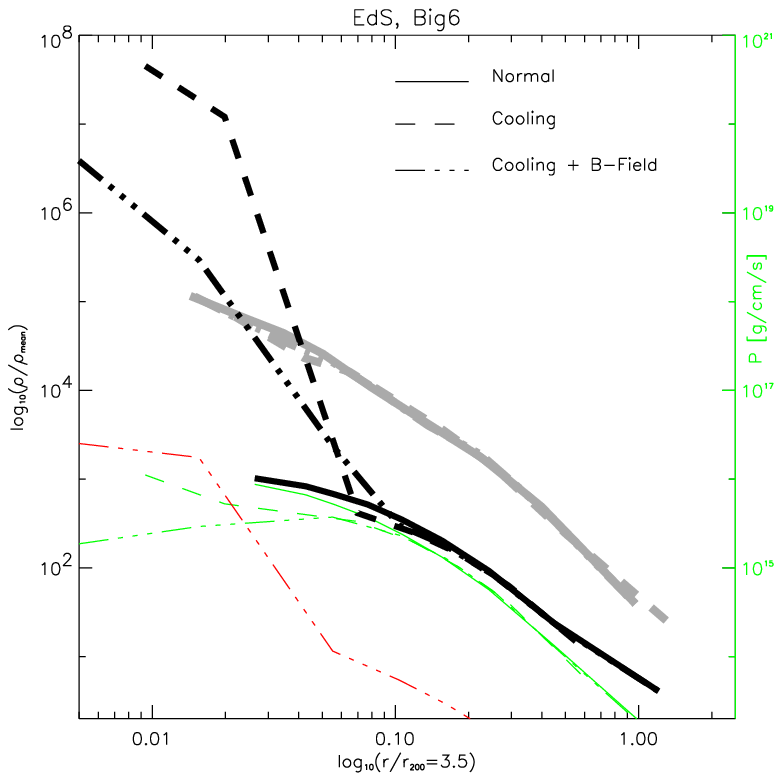,width=0.49\textwidth}
\psfig{figure=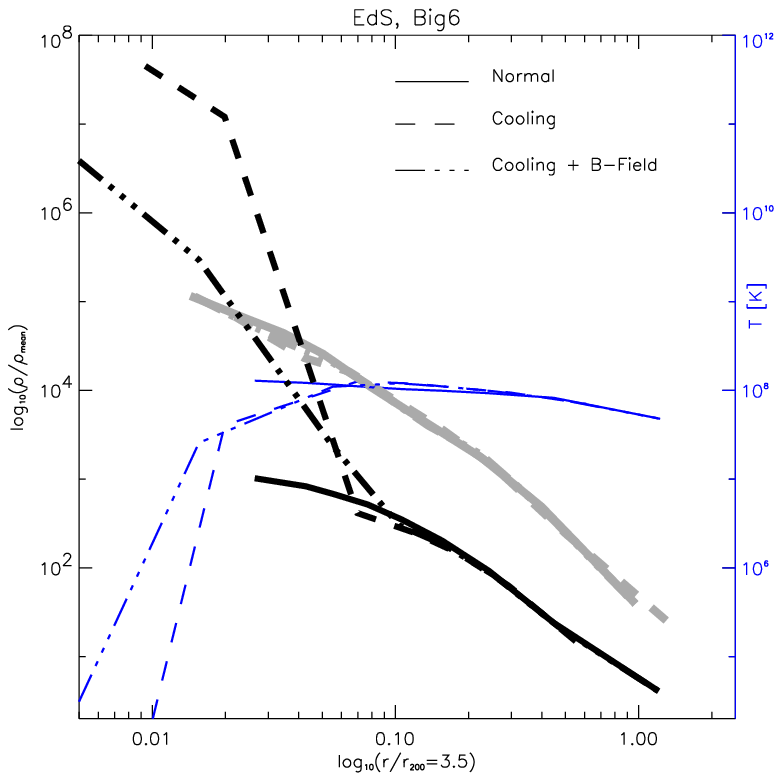,width=0.49\textwidth}
\end{center}
\caption{Shown are the results for one simulated cluster without cooling, with
cooling and with cooling and back-reaction of the magnetic field. Shown
are the density (dark and gray lines) and the temperature (left panel)
or the pressure (right panel). The magnetic pressure (red line)
exceeds the thermal pressure in the cooling flow region.\label{fig:cool}}
\end{figure}

It is known from observations that strong magnetic fields appear in 
cooling-flows. Turning cooling on in our simulations, the collapse of
the cool regions strongly amplifies the magnetic field. The magnetic
field reaches the regime, where the magnetic pressure exceeds the
thermal pressure and stops the collapse of the gas. The synthetic rotation
measures in these cool regions are well in agreement with the
observed values of thousands of rad/m$^2$. For details see Dolag,
Bartelmann \& Lesch (2000b). The conclusions drawn 
for additional processes in our simulations can be summarized as follows.
  
  \begin{itemize}
      \item[{\color{blue}$\bullet$}] The energy content of
          relativistic protons needed to produce enough relativistic
          electrons to get typical radio luminosities for the simulated
          clusters lies between 4\% and 15\% of the thermal energy
          content of the gas (in the range of magnetic
          field strength suggested by Faraday measurements).
      \item[{\color{blue}$\bullet$}] The synthetic radio halo of one
          simulated cluster with comparable mass and temperature
          reproduces the radial profile observed in Coma very well.
      \item[{\color{blue}$\bullet$}] Using one normalization for the
          whole set of simulations the simulation predicts the
          observed, strong correlation between the temperature and the
          radio luminosity of galaxy clusters. 
      \item[{\color{blue}$\bullet$}] For simulations allowing the ICM
          to cool, the magnetic pressure becomes important for the
          dynamics of the regions with strong cooling. The
          temperature drops less and the cool regions get less dense 
          in the presence of magnetic fields. 
      \item[{\color{blue}$\bullet$}] The synthetic Faraday rotation measurements in the
          cooling-flow regions reach the observed extreme values.
   \end{itemize}

\newpage

\section*{References}

  \newcommand{\mybibtitel}[1]{\\{\sl "#1"}\\}


\begin{thebibliography}{99}

\bibitem{DBL99} Dolag, K., Bartelmann, M., Lesch, H., 1999:
         \mybibtitel{SPH simulations of magnetic fields in galaxy clusters}
         \Journal{A\&A}{348}{351}{1999}
\bibitem{S90} Sugimoto, D., Chikada, Y., Makino,J., Ito, T., Ebisuzaki,
            T., Umemura, M., 1990: 
            \mybibtitel{A Special-Purpose Computer for Gravitational Many-Body Problems}
            \Journal{(Nat}{345}{33}{1990}
\bibitem{St96} Steinmetz, M., 1996: 
         \mybibtitel{GRAPESPH: cosmological
            smoothed particle hydrodynamics simulations with the
            special-purpose hardware GRAPE}
          \Journal{MNRAS}{278}{1005}{1996}
\bibitem{R94} Rees, M.J., 1994: 
         \mybibtitel{Origin of sees magnetic field for a galactic dynamo}
         in {\em Cosmic Magnetism}, ed. D. Lyden-Bell (Kluwer Academic
         Publishers, 1994).
\bibitem{K90} Kim, K.T., Kronberg, P.P., Dewdney, P.E., Landecker, T.L., 1990: 
         \mybibtitel{The halo and magnetic field of the Coma cluster of galaxies}
          \Journal{ApJ}{355}{29}{1990}
\bibitem{F99} Ferretti, L., Dallacasa, D., Govoni, F., Giovannini, G.,
         Taylor, G. B.,  Klein, U., 1999: 
         \mybibtitel{The radio galaxies and the magnetic field in Abell 119}
          \Journal{A\&A}{344}{472-482}{1999}
\bibitem{K91} Kim, K.T., Kronberg, P.P., Tribble, P.C., 1991: 
         \mybibtitel{Detection of excess rotation measure due to intracluster magnetic fields in clusters of galaxies}
         \Journal{ApJ}{379}{80}{1991}
\bibitem{DBL00a} Dolag, K., Bartelmann, M., Lesch, H., 2000a:
         \mybibtitel{Evolution and structure of magnetic fields in simulated galaxy clusters}
         In preperation.
\bibitem{DEB00} Dolag, K., Evrard, A., Bartelmann, M., 2000:
         \mybibtitel{The temperature-mass relation in magnetized galaxy clusters}
         Submitted to {\bf A\&A}.
\bibitem{DS00} Dolag, K. \& Schindler, S., 2000:
         \mybibtitel{The effect of magnetic fields on the mass determination of clusters of galaxies}
         Accepted for publication in {\bf A\&A}.
\bibitem{D97} Deiss, B.M., Reich, W., Lesch, H., Wieblebinski, R., 1997:
         \mybibtitel{The large-scale structure of the diffuse radio
            halo of the Coma cluster at 1.4 GHz}
          \Journal{A\&A}{321}{55-63}{1997}
\bibitem{DE00} Dolag, K. \& Ensslin, T., 2000:
         \mybibtitel{Radio Halos of Galaxy Clusters from Hadronic Secondary Electron Injection in
            Realistic Magnetic Field Configurations}
            Accepted for publication in {\bf A\&A}.
\bibitem{DBL00b} Dolag, K., Bartelmann, M., Lesch, H., 2000b: \\
         In preperation.



\end{thebibliography}
\end{document}